\documentclass[fleqn]{revtex4}%
\usepackage{amsfonts}
\usepackage[fleqn,nosumlimits,nonamelimits,reqno]{amsmath}
\usepackage{amssymb}
\usepackage{graphicx}%
\providecommand{\U}[1]{\protect\rule{.1in}{.1in}}

\begin{document}
\preprint{preprint}
\title{The impact of local resonance on the enhanced transmission and dispersion of
surface resonances}
\author{Zeyong Wei}
\affiliation{Physics Department,Tongji University,200092,Shanghai,China}
\author{Jinxin Fu}
\affiliation{Physics Department,Tongji University,200092,Shanghai,China}
\author{Yang Cao}
\affiliation{Physics Department,Tongji University,200092,Shanghai,China}
\author{Chao Wu}
\affiliation{Physics Department,Tongji University,200092,Shanghai,China}
\author{Hongqiang Li}
\email{hqlee@tongji.edu.cn}
\affiliation{Physics Department,Tongji University,200092,Shanghai,China}
\pacs{PACS number}

\begin{abstract}
We investigate the enhanced microwave transmission through the array of
metallic coaxial annular apertures (MCAAs) experimentally and theoretically.
The even-mode and the odd-mode surface resonances are clarified from the
spatial field distributions and the dispersion diagram. The impact of local
resonance is thoroughly embodied in the even-mode surface resonant states,
while the odd-mode surface resonances are scaled by periodicity, invariant to
different local geometry of the unit cell, and invisible in measurements. The
enhanced transmission is the collective selections on the interplay between
the local resonances and the evanescent Bloch wave channels on the surface.
Transmission measurements for different inner diameter of the apertures show
that the transmissivity extrema with respect to the specific angles precisely
correspond to the degenerate points in the dispersion diagram of surface resonances.

\end{abstract}
\maketitle

\section{INTRODUCTION}

Extraordinary optical transmission (EOT) through a periodical array of
subwavelength nano-holes in metal films\cite{1} provides a new way to
manipulate the light waves in subwavelength division. The origin of EOT was
attributed to the resonant tunneling of surface plasmon polaritons
(SPPs)\cite{2}. Later studies found that the enhanced transmission can also be
achieved in Terahertz\cite{3} or microwave region\cite{4} where the metal
surface does not support SPPs but surface resonances\cite{5}.\emph{ }The high
transmission peak through more complicated apertures, such as coaxial
apertures\cite{6,7}, rectangular holes\cite{8,9}, etc., tends to be
red-shifted or even scaled by waveguide resonance instead of the
periodicity\cite{10}.\emph{ }Polarization-dependent transmission spectra as
well as the optical activities are observed when the array of large-sized
elliptal holes is adopted\cite{11}. These studies strongly indicate that the
local resonance is crucial and dominant to the formation of EOT under specific circumstances.

In this paper, we adopt the modal expansion method (MEM) to calculate the
enhanced microwave transmission through the array of metallic coaxial annular
apertures (MCAAs) as well as the dispersion of surface resonances. In
microwave region, metals are perfect electric conductors (PECs), which\emph{
}enable us to easily characterize the spatial dispersion from the periodic
structure. The even-mode and the odd-mode surface resonances are clarified by
calculations on the transmission spectra, the spatial field distributions, and
the dispersion diagram. The impact of local resonance is thoroughly embodied
in the even-mode surface resonant states, while the odd-mode surface
resonances are scaled by periodicity, invariant to different local geometry of
the unit cell, and invisible in measurements. And the enhanced transmission
comes from collective selections on the interplay between the local resonances
and the evanescent Bloch wave channels on the surface. Transmissivity extrema
with respect to the specific incident angles precisely correspond to the
degenerate points in the dispersion diagram of surface resonances. The results
are heuristic to the enhanced transmission in the optical region as well.

\section{SAMPLE DESCRIPTION AND EXPERIMENTAL SETUP}

Our samples were fabricated on $1$m$\times1$m copper slabs with a thickness of
$t=2$mm. As a first step to fabricate our samples, two-dimensional square
arrays of air holes with a lattice constant $p=30$mm were drilled through four
identical copper plates, each hole has the same diameter $D=10$mm; Secondly,
each air hole was embeded coaxially with an annular ring made of teflon with a
dielectric constant $\varepsilon_{r}=2.2$. and an appropriate diameter $d$;
Finally, a copper cylinder with a diameter of d is embeded in the dielectric
ring at the hole center. as shown in the inset of Fig.1. Four samples were
fabricated with different inner diameters $d=8.3$, $6.3$, $5.0$, $2.0$mm respectively.

The transsmission measurements are performed with Aglient network analyzer
8722ES in a microwave chamber. Three pairs of highly directive horn antennas
with a gain factor of 24.8dB are utilized to cover the frequency range within
5.8$\sim$8.2, 8.2$\sim$12.4, and12.4$\sim$18.2GHz respectively. The sample
plate is vertically positioned on a rotary table, at the center of the line
between the emitting and the receiving horns, the horn antennas are 5.0 meters
apart from each other. Oblique incidence is achieved by rotating the sample
plate axially about $y$ axis. Horizontal($\vec{E}||\hat{x}$) or vertical
polarized ($\vec{E}||\hat{y}$) Gaussian beams from the horns give rise to the
$TM$ or $TE$ polarized incidence. Transmissivity throught the sample plate is
normalized to the wave energy transmitted between the two horn antennas in
free space without the sample plate.

\section{MODAL EXPANSION THEROY}

The modal expansion method\cite{12} is a powerful tool to calculate the
transmission of layered system, for example, one-dimentional metallic grating.
The essence of the method is that the electromagnetic wave fields in the
specific layers can be decomposed with a series of in-plane eigen-functions
correspondingly. The problem can be solved by applying boundary continuum
condition at all interfaces. At low frequency region, the wave fields inside
the aperture mainly determined by the fundamental waveguide mode, so that the
method can be simplified to one-mode theory\cite{13} which is fast and convergent.

With the notions stated above, we develop the modal expansion method to deal
with two-dimentional(2D) array of MCAAs. We note that the plane wave incidence
can not excite $TEM$ guided mode and all the $TM$ guided modes in a coaxial
aperture accounting for its in-plane symmetry. So we only consider the
$TE_{p,q}$ guided mode inside the MCAAs, where the integer pair $(p,q)$
denotes the mode number\cite{14}. The magnetic field inside the MCAAs within
the region of $0<z<h$ can be written as the superposition of all the forward
and backward $TE_{p,q}$ modes:%
\begin{equation}
\vec{H}^{II}=\sum_{p=1}^{\infty}\sum_{q=1}^{\infty}\left(  a_{p,q}%
e^{-i\beta_{p,q}z}-b_{p,q}e^{i\beta_{p,q}z}\right)  \beta_{p,q}\vec{g}%
_{p,q}(x,y), \label{eq_h2}%
\end{equation}
where $a_{p,q}$ and $b_{p,q}$ are the coefficents of the $(p,q)^{th}$ order of
forward and backward guided waves, $\beta_{p,q}$ is the wavevector along $z$
axis with $\beta_{p,q}=\sqrt{k^{2}-T_{p,q}^{2}}$,and $T_{p,q}$ is the $q^{th}$
root of eigen function $J_{p}^{\prime}\left(  Td/2\right)  N_{p}^{\prime
}\left(  TD/2\right)  -J_{p}^{\prime}\left(  TD/2\right)  N_{p}^{\prime
}\left(  Td/2\right)  =0$. $\vec{g}_{p,q}(x,y)$ is the in-plane mode profile
of the $(p,q)^{th}$ order of guided wave, which can be expressed with $\vec
{g}_{p,q}(x,y)=g_{\rho_{p,q}}(\rho,\phi)\hat{e}_{\rho}+g_{\phi_{p,q}}%
(\rho,\phi)\hat{e}_{\phi}$ in cylinder coordinates, where%

\begin{align}
g_{\rho_{p,q}}(\rho,\phi)  &  =\left[  N_{p}^{\prime}\left(  T_{p,q}%
d/2\right)  J_{p}^{\prime}\left(  T_{p,q}\rho\right)  -J_{p}^{\prime}\left(
T_{p,q}d/2\right)  N_{p}^{\prime}\left(  T_{p,q}\rho\right)  \right]
\sin(p\phi)\nonumber\\
g_{\phi_{p,q}}(\rho,\phi)  &  =-\frac{p}{\rho}\left[  N_{p}^{\prime}\left(
T_{p,q}d/2\right)  J_{p}\left(  T_{p,q}\rho\right)  -J_{p}^{\prime}\left(
T_{p,q}d/2\right)  N_{p}\left(  T_{p,q}\rho\right)  \right]  \cos(p\phi)
\label{eq_gl}%
\end{align}
The wave fields at the incident side($z<0$) and the outgoing side ($z>h$)of
the slab can be expanded as the superposition of all orders of Bloch waves respectively%

\begin{equation}
\vec{H}^{I}=I_{0,0}e^{-i\vec{k}_{\parallel}\cdot\vec{r}_{\parallel}%
}e^{-ik_{z_{0,0}}z}+\sum_{m=-\infty}^{+\infty}\sum_{n=-\infty}^{+\infty
}R_{m,n}e^{-i(\vec{k}_{\parallel}+\vec{G}_{m,n})\cdot\vec{r}_{\parallel}%
}e^{-ik_{z_{m,n}}z} \label{eq_h1}%
\end{equation}%
\begin{equation}
\vec{H}^{III}=\sum_{m=-\infty}^{+\infty}\sum_{n=-\infty}^{+\infty}%
T_{m,n}e^{-i(\vec{k}_{\parallel}+\vec{G}_{m,n})\cdot\vec{r}_{\parallel}%
}e^{-ik_{z_{m,n}}z} \label{eq_h3}%
\end{equation}
where $I_{0,0}$ represents the coefficent of incident wave, $\vec
{k}_{\parallel}=k_{0}\sin(\theta)\hat{e}_{x}$ is the in-plane wavevector,
while $k_{z_{0,0}}=\sqrt{k_{0}^{2}-|\vec{k}_{\parallel}|^{2}}$ is the
wavevector along $z$ axis. $R_{m,n}$ and $T_{m,n}$ are the coefficents of the
$(m,n)^{th}$ reflected and transmitted Bloch waves. $\vec{G}_{m,n}=\hat{e}%
_{x}\frac{2\pi m}{p}+\hat{e}_{y}\frac{2\pi n}{p}$ is the Bloch wavevector of
square lattice, where $\hat{e}_{x}$and $\hat{e}_{y}$are the unit vectors of
reciprical space in cartesian coordinate representations. After matching the
wave fields of the three regions at two interfaces $z=0$and$z=h$, we have
\begin{align}
\sum_{m=-\infty}^{\infty}\sum_{n=-\infty}^{\infty}\Omega_{m,n,p,q}%
(I_{0,0}\delta_{m}\delta_{n}+R_{m,n})  &  =a_{p,q}+b_{p,q}\nonumber\\
\frac{k_{z_{m,n}}}{\varepsilon_{0}}(I_{0,0}\delta_{m}\delta_{n}-R_{m,n})  &
=\sum_{p=1}^{\infty}\sum_{q=1}^{\infty}\frac{\Omega_{m,n,p,q}^{\ast}%
\beta_{p,q}}{\varepsilon_{r}}\left(  a_{p,q}+b_{p,q}\right) \nonumber\\
\sum_{m=-\infty}^{\infty}\sum_{n=-\infty}^{\infty}\Omega_{m,n,p,q}T_{m,n}  &
=a_{p,q}e^{-i\beta_{p,q}h}+b_{p,q}e^{i\beta_{p,q}h}\nonumber\\
\frac{k_{z_{m,n}}}{\varepsilon_{0}}T_{m,n}  &  =\sum_{p=1}^{\infty}\sum
_{q=1}^{\infty}\frac{\Omega_{m,n,p,q}^{\ast}\beta_{p,q}}{\varepsilon_{r}%
}[a_{p,q}e^{-i\beta_{p,q}h}-b_{p,q}e^{i\beta_{p,q}h}] \label{eq_eqs}%
\end{align}
where $\Omega_{m,n,p,q}=\int_{0}^{2\pi}d\phi\int_{a}^{b}d\rho e^{i(\vec
{k}_{\parallel}+\vec{G}_{m,n})\cdot\vec{r}_{\parallel}}\vec{g}_{l}^{\ast}%
(\rho,\phi)$ is an overlap intergral, denoting the coupling coefficent from a
Bloch wave channel $(m,n)^{th}$ to a guilded wave channel $(p,q)^{th}$. We
expand a plane wave $e^{-ik_{\parallel}\rho\cos\left(  \phi-\theta\right)  }$
in the form of $\sum_{l=0}^{\infty}\frac{1}{l!}\left[  -ik_{\parallel}\rho
\cos\left(  \phi-\theta\right)  \right]  ^{l}$ in cylinder coordinates to
solve EQs(\ref{eq_eqs}). Then we derived coefficents of the transmitted or
reflected Bloch waves for the linear EQs(\ref{eq_eqs})%

\begin{align}
T_{m,n}  &  =\Omega_{m,n,p,q}^{\ast}\frac{k_{z_{m,n}}}{\beta_{p,q}}%
(a_{p,q}e^{-i\beta_{p,q}h}-b_{p,q}e^{i\beta_{p,q}h})\nonumber\\
R_{m,n}  &  =I_{0,0}\delta_{m}\delta_{n}-\Omega_{m,n,p,q}\frac{k_{z_{m,n}}%
}{\beta_{p,q}}(a_{p,q}-b_{p,q}) \label{eq-coeff}%
\end{align}
As surface resoances are intrinsic properties of the system, we can extract
the dispersion of surface resonances by assigning a zero value to the
incidence, i.e., omitting the term $I_{0,0}e^{i\vec{k}_{\parallel}\cdot\vec
{r}_{\parallel}}e^{-ik_{z_{0,0}}z}$in EQ(\ref{eq_h1}), and solving the
eigen-value of EQs(\ref{eq_eqs}).

\section{EVEN-MODE OR ODD-MODE SURFACE RESONANCES}

Circular dots and solid lines in Fig.1) presented the measured and calculated
transmission spectra under normal incidence ($\theta=0^{\text{o}}$) for
samples with diffferent inner diameters $d=8.3,6.3,5.0,2.0$mm. For the sample
with $d=8.3$mm, the measured transmission peak reaches unity at $8.64$GHz
($Fig.1a)$). When the inner diameter $d$ becomes smaller, the transmission
peak at $8.64GHz$ becomes blue-shifted. In good agreement with the peak and
lineshape of the measured curves, the calculated transmission spectra also
possess two narrow transmission peaks with fano lineshape at $9.99$GHz and
$14.12$GHz irrespective to\emph{ }the inner diameter $d$. These additional
transmission peaks are scaled by periodicity of the structure. Fig.2) presents
frequencies of the transmission peaks under normal incidence ($\theta
=0^{\text{o}}$)with respect to different inner diameters $d$. We calculate the
spatial field distributions on the slab to better understand the formation and
the underlying mechanism of these two kinds of transmission peaks. $Fig3)$
presents the electric field component and the power flow component in $xz$
plane on the slab surface with respect to the two transmission peak at
$8.64$GHz and at $9.99$GHz. For the narrow resonant peak at $9.99$GHz,
calculations on the spatial field distribution show that the electric field
component along z axis is anti-symmetric to the $z=\frac{h}{2}$ plane at both
sides of the plate, which is rightly the field distribution of an odd-mode
surface resonance \cite{15}, as shown in $Fig.3b)$ and $Fig.3d)$. It is
reasonable that its frequency is mainly decided by the periodicity and
symmetry of the array, and indifferent to the local geometry of apertures, in
that most field energy near the metallic surface distributes far from the
apertures with the minimum in them, as shown in Fig.3d). In other words, the
odd-mode surface resonance is almost blind to the aperture. While the shifted
broad transmission peaks in $Fig.2)$ come from the even-mode surface resonant
states in that the electric field component along z axis is symmetric to the
$z=\frac{h}{2}$ plane at both sides of the plate with most field energy
localized inside the aperture as well as its vicinity near the surface, as
shown in $Fig.3a)$ and $Fig.3c)$ for the transmission peak at $8.64GHz$
respectively. The normalized strength of the electric field inside the
aperture can be enhanced to 1\symbol{126}2 orders depending on the straitness
of the radial gap width $(D-d)/2$. The field enhancement may be applied to the
near-field probing or nonlinear optics \cite{16}. The frequency of an odd-mode
surface resonance $f_{m,n}^{Odd}(k_{x})$ is slightly lower than Rayleigh
frequency $f_{m,n}^{Rayleigh}(k_{//}^{^{\prime}})=\frac{c}{p}\sqrt
{(k_{x}^{\prime}+m)^{2}+(k_{y}^{\prime}+n)^{2}}$where Wood anomaly happens
\cite{17,18}, $k_{//}^{^{\prime}}=(k_{x},k_{y})\cdot p/(2\pi)$is the
normalized wavevector. For example, at $k_{x}=k_{y}=0$, $f_{1,0}%
^{Rayleigh}\simeq9.993GHz$ is slightly higher than $f_{1,0}^{Odd}\simeq
9.990$GHz\ for the array of holes ($d=8.3mm$). With the increase of $d$, the
frequency $f_{m,n}^{Odd}(k_{//}^{^{\prime}})$ suffers an imperceptible
blue-shift. All the even-mode branches in $Fig.2)$ are confined within the
intervals devided by the frequencies \{$f_{m,n}^{Odd}$\} of odd-mode surface
resonances, and tends to asymptotic to the lower bound with the increase
of\ $d$. The features imply that the even-mode surface resonances are the
collective selections on the interplay between the waveguide resonance and
in-plane evanescent diffractive channels.

\section{Angular dependent transmission spectra and dispersion diagram of
surface resonances}

We calculate the dispersion diagram of surface resonances for all
polarizations, as shown in the solid or dashed lines in Fig.4) for the
even-mode or odd-mode surface resonant states.We also measured the
transmission spectra through all four samples under oblique incidence. The
mesured transmissivity is plotted in colormap as a function of frequency and
the in-plane wavevector $\vec{k}_{\parallel}=\frac{2\pi f}{c_{0}}\sin
(\theta)\hat{e}_{x}$,\emph{ }converted from the incident angle $\theta$. For
the sample of $d=8.3$mm, a transmission peak at $f=8.64$GHz in $TE$
polarization is not shifed almost under any incidence angles (see $Fig.4a)$),
reaching unity at normal incidence\ with $\theta=0^{\text{o}}$. A similar
result has been predicted in a previous study on the rectangular
holes\cite{10}. For the other three samples, the first transmission peak in
$TE$ polarization become dispersive alternatively with respect to the incident
angles, the experimental data for $d=6.3,5.0,2.0$mm is shown in $Fig.4c$,
$4e$, $4g)$ respectively. It implies that the local resonance of the sample
with $d=8.3$mm lies below the limit of Rayleigh frequency while above it for
other samples, as we know that a local resonance will couple with different
Bloch channels when it is below or above Rayleigh frequency.

Below the limit of Rayleigh frequency\ $f_{1,0}^{Rayleigh}$, there only
exist\ $\vec{G}(-1,0)$\ and $\vec{G}(0,0)$ Bloch channels. Without the
projection of wave vector component of the incidence along $y$ axis, the $TE$
polarized waves incident in the $y=0$ plane can not couple with these Bloch
channels. So for $d=8.3$mm, without the assistance from any Bloch channels, we
can observe a flat polariton branch in $TE$ mode, as shown in Fig.4a).
Meanwhile the TM polarized incidence in the $y=0$ plane can be coupled to the
$\vec{G}(-1,0)$ Bloch channel due to the same reason; the $TM$ poliarized
surface resonant states in even mode comes from the interference between the
waveguide resonance and the $\vec{G}(-1,0)$ Bloch channels, giving rise to two
anti-symmetric asymptotic branches in $Fig.4b)$. The bright curvatures for
measured data prove the existance of the two branches, as shown $Fig.4b)$.
Thus the role of local resonance on the surface waves below $f_{1,0}%
^{Rayleigh}$ is revealed.

More diffractive channels lie above $f_{1,0}^{Rayleigh}$, such as $\vec
{G}(0,\pm1)$, and$\vec{G}(\pm1,\pm1)$ etc. When $d$ becomes smaller, the
waveguide resonance will shift across $f_{1,0}^{Rayleigh}$ from below. $TE$
polarized incidence will couple with the waveguide resonance via $\vec
{G}(0,\pm1)$, and$\vec{G}(-1,\pm1)$ channels, resulting in different features
of dispersion diagrams and transmission spectra. As an example, for the sample
with $d=6.3$mm, the three branches below $f_{1,1}^{Odd}=14.12$GHz are rightly
the consequence of the interplay between the local resonance and the $\vec
{G}(0,\pm1)$, $\vec{G}(-1,\pm1)$ channels under $TE$ polarized incidence. As
shown in Fig.4c), the bright-colored curvature exactly fits the curve of the
lowest branch with the brightest point at the location ($k_{x}=\pi/p$,
$f=9.97$GHz), which is the four-fold degenerate point about $\vec{G}(-1,\pm
1)$, and $\vec{G}(0,\pm1)$. The maximal transmissivity at the Brillouin zone
boundary means that the strongest coupling between the local resonance and the
Bloch channels is likely to occur at the degenrate states with high symmetry.
The coupling between local resonance and diffractive channels will induce
highly inhomogeneous field distribution on the surface, resulting in the
opening of bandgaps, similar to the case of corrugated metallic surface
\cite{19}. As shown in Fig.4c), existence of two bandgaps between the three
branches verifies the analysis. The coupling will inevitably induce the split
of the frequency at the degenrate points. We notice that right above the
\textquotedblleft brightest\textquotedblright\ state ($k_{x}=\pi/p$,
$f=9.97$GHz) , the second $TE$ polarized branch in even mode converge with
four odd-mode branches $(-1,\pm1)$ and $(0,\pm1)$ at the degenerate
state($k_{x}=\pi/p$, $f=$11.12GHz) on BZ boundary. The degeneracy demonstrates
that the state also possesses anti-symmetric distribution of the spatial field
and minimum transmissivity. It also applies for Fig.4e) and Fig.4g); the
superposed bright curvature on the second branch becomes weaker, reaching
minimum at the four-fold degenerate state. The rule governs all dispersion
diagrams in TM polarization, on the second branch at $k_{x}=\pi/p$, the third
branch at $k_{x}=\pi/2p$ and $k_{x}=3\pi/2p$, the fourth branch at $k_{x}%
=\pi/2p$ in Fig.4b,d,f,h). Transmission extrema at the degenerate states are
understandable. Considering these brightest states with four-fold degeneracy,
they are likely to couple with the incident waves with strongly localized
field within the apertures and have zero group velocity along any direction
within the $xy$ plane. Meanwhile, the odd-mode surface resonant states are
invisible in measurements, and the geometry of apertures has little impact on
the odd-mode branches.

\section{CONCLUSION}

In summary, we perform microwave experiments and rigorous modal expansion
method to investigate the enhanced microwave transmission through the array of
metallic coaxial annular apertures (MCAAs) as well as the dispersion of
surface resonances. The even-mode and the odd-mode surface resonances are
clarified theoretically by the spatial field distributions and the dispersion
diagram. The impact of local resonance is thoroughly embodied in the even-mode
surface resonant states. While the odd-mode surface resonant states, invisible
in measurements, are scaled by periodicity, invariant to different local
geometry of the unit cell. The enhanced transmission is the collective
selections on the interplay between the local resonances and the evanescent
Bloch channels on the surface. Transmissivity reaches extrema at the
degenerate points in the dispersion diagram of surface resonances. The work
provides a holisitc description on the interplay between the local resonance
and the plane wave excitations via evanesent Bloch channels.

\begin{acknowledgments}
This work was supported by the National 863 Program of China (Grant
No.2006AA03Z407), NSFC (Grant No.10574099, No.60674778), CNKBRSF(Grant
No.2006CB921701), NECT, STCSM and Shanghai Education and Development
Foundation (No. 06SG24).
\end{acknowledgments}

\newpage

{\huge Figure captions:}\newline

FIG.1.Simulated(lines) and Measured(open circles) transmission spectra through
the array of MCAAs with different inner diameter $d=8.3$mm(a),$6.3$mm(b),
$5.0$mm(c), $2.0$mm(d) respectively. The inset presents the photo of the
sample ($d=8.3$mm)\newline

FIG.2.Resonant frequencies of even mode(solid lines) and odd mode(dashed
lines) under normal incidence ($k_{\parallel}=0$) with respect to the inner
diameter $d$ varying in the range of $0.0\sim4.9$mm. The measured frequencies
for different samples are marked with stars.\newline

FIG.3. Spatial field distributions of the even/odd surface mode in $xz$ plane
for the sample with $d=8.3$mm. (a) E field, even; (b) E field, odd; (c) Power
flow, even; (d)Power flow, odd\newline

FIG.4. Dispersion relations of even surface resonances (solid lines) and odd
surface surface resonances (dashed lines) for samples with different inner
diameter $d=8.3$, $6.3$, $5.0$, and $2.0$mm. (a), ( c), (e), (g) for $TE$
polarization and (b), (d),( f), (h) for $TM$ polarization. Frequency $f$ is
normalized to the limit of Rayleigh frequency $c0/p=9.993$GHz. Measured
transmittivities with respective to the in-plane wave vector and frequency are
plotted in color-scale.\newpage

\begin{center}%
\begin{center}
\includegraphics[
height=3.0364in,
width=3.96in
]%
{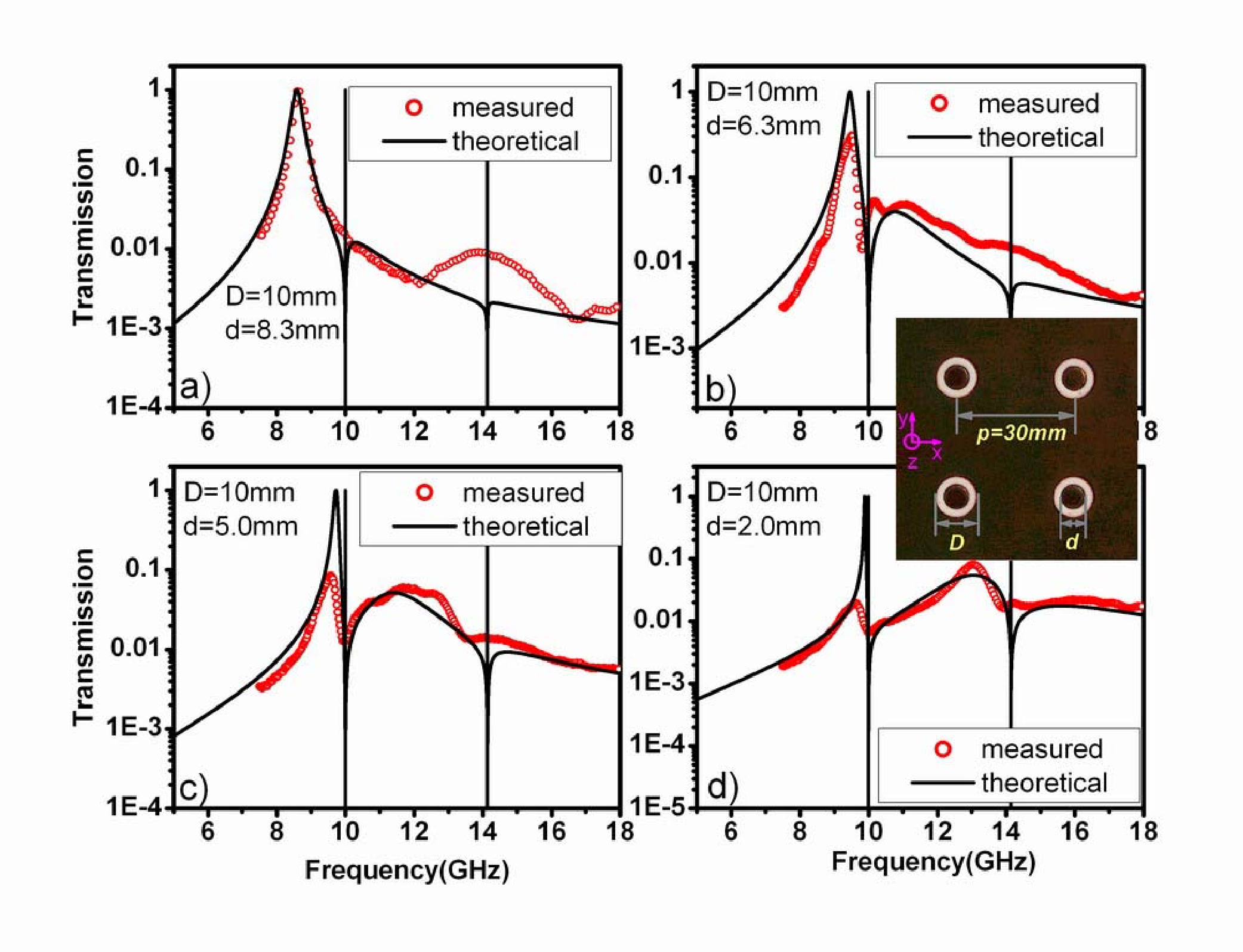}%
\\
FIG.1.Simulated(lines) and Measured(open circles) transmission spectra through
the array of MCAAs with different inner diameter $d=8.3$mm(a),$6.3$mm(b),
$5.0$mm(c), $2.0$mm(d) respectively. The inset presents the photo of the
sample ($d=8.3$mm)
\end{center}
%

\begin{center}
\includegraphics[
height=3.0364in,
width=3.9643in
]%
{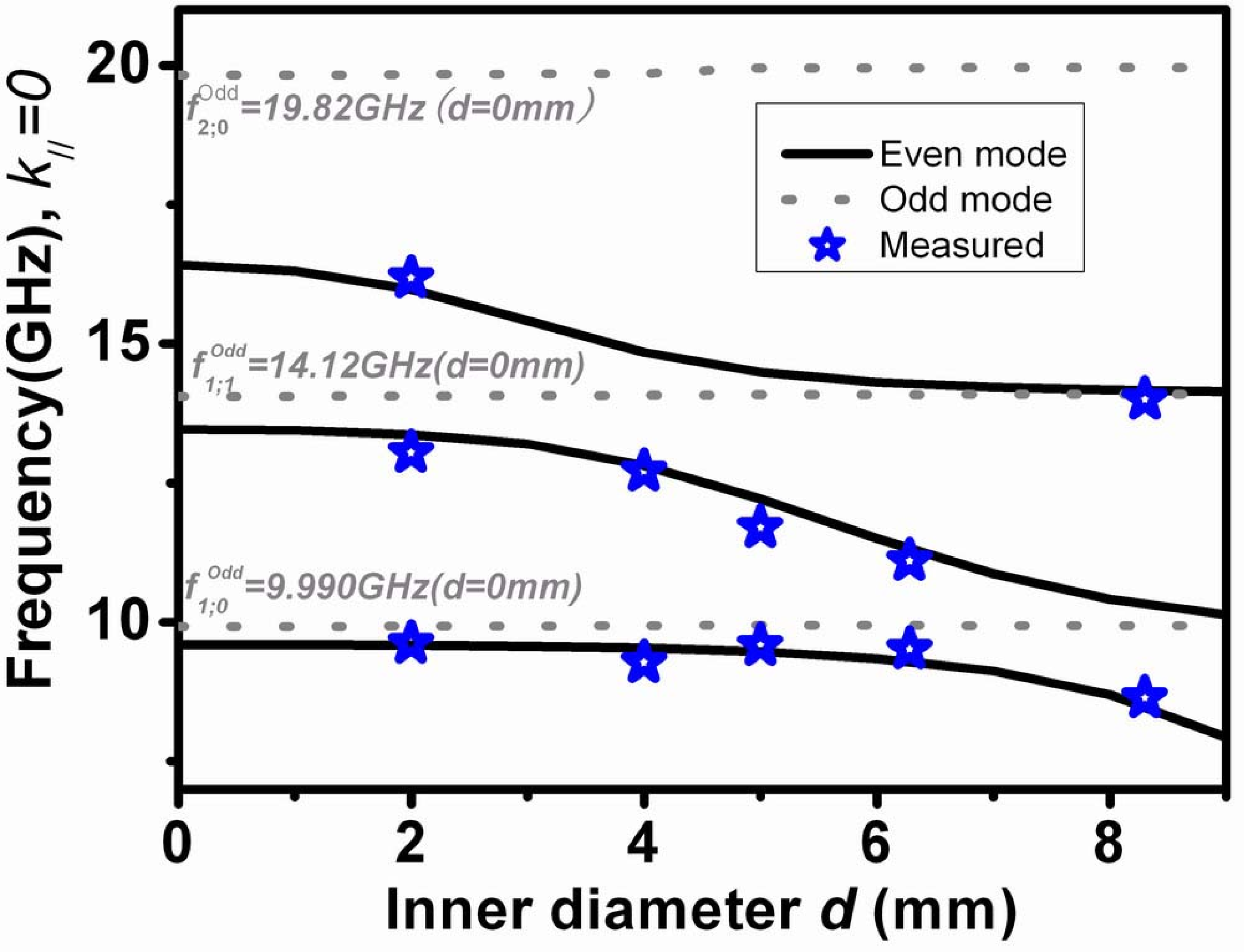}%
\\
FIG.2.Resonant frequencies of even mode(solid lines) and odd mode(dashed
lines) under normal incidence ($k_{\parallel}=0$) with respect to the inner
diameter $d$ varying in the range of $0.0\sim4.9$mm. The measured frequencies
for different samples are marked with stars.
\end{center}
%

\begin{center}
\includegraphics[
height=3.0355in,
width=3.0485in
]%
{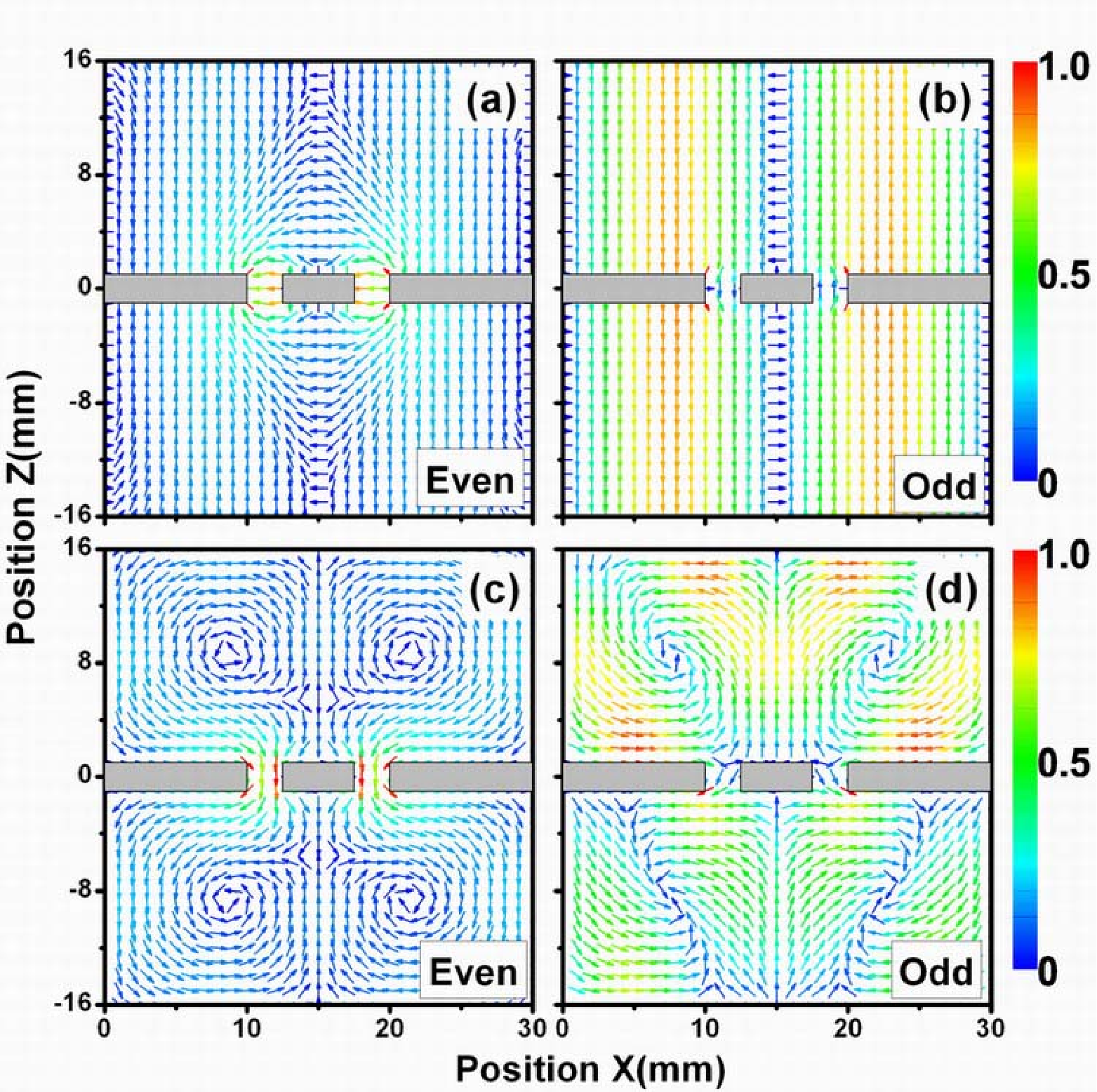}%
\\
FIG.3. Spatial field distributions of the even/odd surface mode in $xz$ plane
for the sample with $d=8.3$mm. (a) E field, even; (b) E field, odd; (c) Power
flow, even; (d)Power flow, odd
\end{center}
%

\begin{center}
\includegraphics[
height=3.0364in,
width=2.5538in
]%
{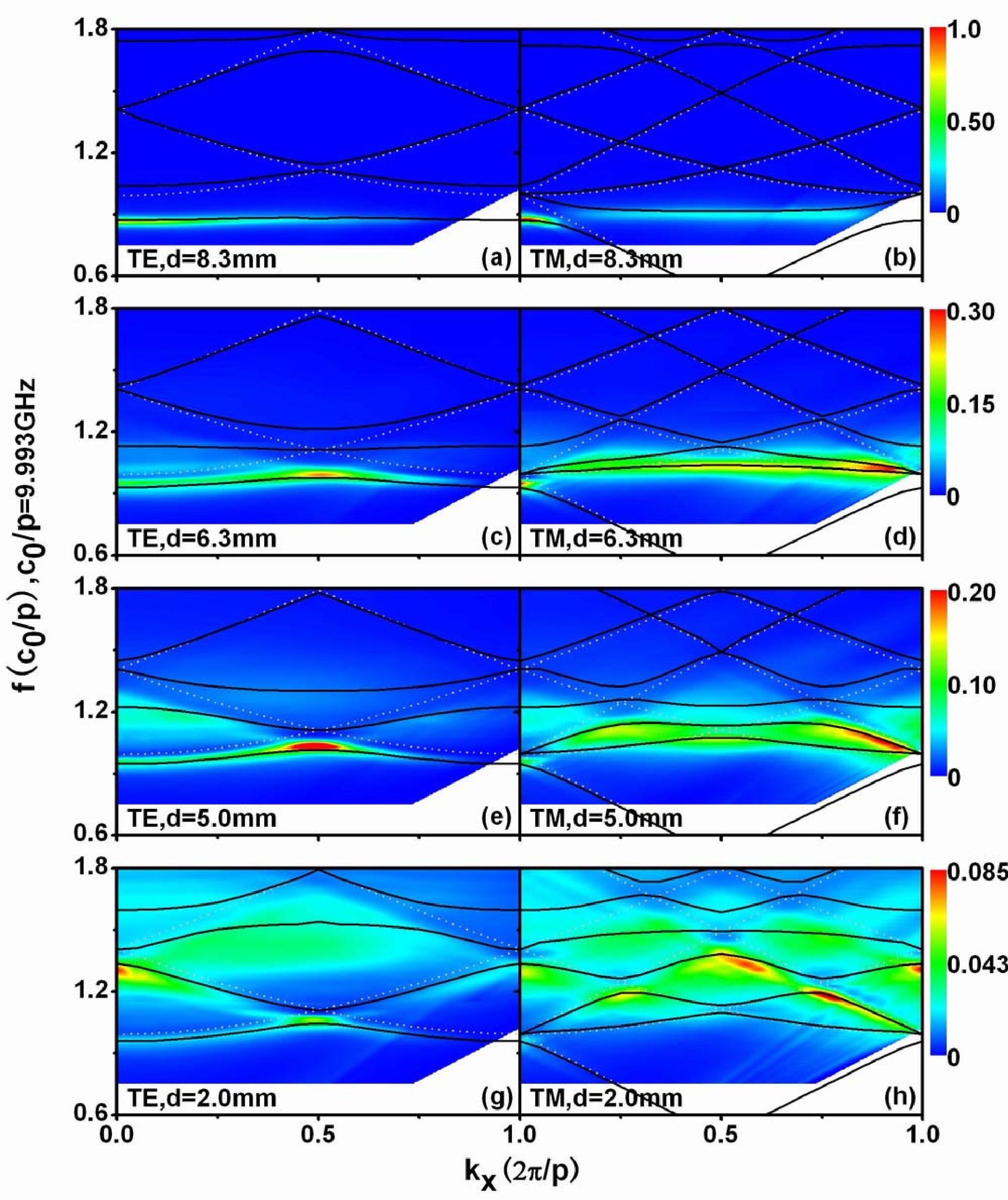}%
\\
FIG.4. Dispersion relations of even surface resonances (solid lines) and odd
surface surface resonances (dashed lines) for samples with different inner
diameter $d=8.3$, $6.3$, $5.0$, and $2.0$mm. (a), ( c), (e), (g) for $TE$
polarization and (b), (d),( f), (h) for $TM$ polarization. Frequency $f$ is
normalized to the limit of Rayleigh frequency $c0/p=9.993$GHz. Measured
transmittivities with respective to the in-plane wave vector and frequency are
plotted in color-scale.
\end{center}

\end{center}

\end{document}